\documentclass[12pt,preprint]{aastex}
\shorttitle{Dust envelope of the PPN IRAS19475+3119}
\shortauthors{Sarkar \& Sahai}
\begin{document}
\newcommand{\kms}{\mbox{km~s$^{-1}$}}
\newcommand{\s}{\mbox{$''$}}
\newcommand{\mloss}{\mbox{$\dot{M}$}}
\newcommand{\my}{\mbox{$M_{\odot}$~yr$^{-1}$}}
\newcommand{\ls}{\mbox{$L_{\odot}$}}
\newcommand{\ms}{\mbox{$M_{\odot}$}}
\newcommand\mdot{$\dot{M}  $}
\title{The dust envelope of the pre-planetary nebula IRAS19475+3119}
\author{Geetanjali Sarkar\altaffilmark{1} and Raghvendra
Sahai\altaffilmark{2}}
\affil{Department of Physics, Indian Institute of Technology,
Kanpur-208016, U.P.,
India}
\affil{Jet Propulsion Laboratory, MS 183-900, 4800 Oak Grove Drive,
Pasadena, CA
91109, U.S.A.}
\email{gsarkar@iitk.ac.in, raghvendra.sahai@jpl.nasa.gov}
\begin{abstract}
We present the spectral energy distribution (SED) of the pre-planetary
nebula,
IRAS 19475+3119 (I\,19475), from the optical to the far-infrared. We
identify
emission features due to crystalline silicates in the ISO SWS spectra of
the
star. We have fitted the SED of I\,19475 using a 1-D radiative transfer
code, and
find that a shell with inner and outer radii of 8.8$\times 10^{16}$ and 4.4
$\times 10^{17}$cm, and dust temperatures ranging from about 94\,K to 46\,K
provide the best fit. The mass of this shell is $\gtrsim$1\,[34 cm$^2$
g$^{-1}$/$\kappa(100\mu$m)][$\delta$/200]\ms, where $\kappa$(100$\mu$m) is
the
100$\mu$m~dust mass absorption coefficient (per unit dust mass), and
$\delta$ is
the gas-to-dust ratio. In agreement with results from optical imaging and
millimeter-wave observations of CO emission of I\,19475, our model fits
support
an r$^{-3}$ density law for its dust shell, with important implications for
the
interaction process between the fast collimated post-AGB winds and the
dense AGB
envelopes which results in the observed shapes of PPNs and PNs. We find
that the
observed JCMT flux at sub-millimeter wavelengths (850\micron) is a factor
$\sim$2
larger than our model flux, suggesting the presence of large dust grains in
the
dust shell of I\,19475 which are not accounted for by our adopted standard
MRN
grain size distribution.

\end{abstract}

\section{Introduction}

Pre-Planetary nebulae (PPNs), objects in transition between the AGB and
planetary
nebula (PN) evolutionary phases, hold the key to our understanding of the
relationship between these late evolutionary stages of low and intermediate
mass
($\sim\,1-8$\,\ms) stars. The hydrodynamic interaction of one or more fast,
collimated post-AGB outflows, believed to occur during the PPN phase, with
the
dense, massive, slowly-expanding spherical AGB wind is believed to be
responsible
for shaping planetary nebulae (Sahai \& Trauger 1998, Sahai 2002). Imaging
surveys
with the Hubble Space Telescope
have been crucial in revealing a wide variety of aspherical morphologies in
PPNs
(e.g. Sahai 2004; Ueta et al. 2000),
however the bulk of the circumstellar mass, which often resides in a
spherical,
dusty component surrounding the central aspherical nebula, is not
well-probed in
these data. Thus a crucial ingredient for theoretical studies of such
interactions, namely the mass and density of the
ambient circumstellar medium (e.g., Lee \& Sahai 2003, 2004), is lacking
for a
majority of the PPNs discovered in
HST surveys. Although attempts have been made to estimate the mass from
submillimeter data (e.g. Gledhill et al. 2002), these are compromised by
the
simplifying assumptions about the dust temperature.

The mineralogy of circumstellar dust shells around PPNs became possible
with the wavelength coverage and spectral resolution offered by the
Infrared Space Observatory (ISO) (see e.g. Hrivnak et al. 2000; Molster et
al. 2002a, b, c; Gauba \& Parthasarathy 2004 and references therein). In
order to understand the grain temperatures, mass-loss history and
morphology responsible for the formation of these minerals in the
circumstellar environment of PPNs, detailed modelling of the spectral
energy distributions (SEDs) of a few PPNs (e.g., HD161796, Hoogzaad et al.
2002; IRAS 16342-3814, Dijkstra et al. 2003a; IRAS 16279-4757, Matsuura et
al. 2004; Gauba \& Parthasarathy 2004) have been undertaken.

In this paper, we examine the composition and distribution of dust in the
PPN, IRAS\,19475+3119 (hereafter I\,19475) and model its SED from the
optical to the far-infrared to set constraints on the physical properties
of its dust shell (mass, temperature, size). I\,19475 is listed as a F3Ia
star in the Case-Hamburg luminous stars survey (Stock et al. 1960). High
resolution optical spectra of the star have confirmed its post-AGB nature
(Klochkova et al. 2002; Sivarani et al. 2001; Arellano Ferro et al. 2001).
A point-symmetric nebula with an extent of 4.9\arcsec $\times$
3.4\arcsec~has been seen in J-band polarized flux images (Gledhill et al.
2001). CO J=1--0 and 2--1 emission typical of AGB and post-AGB objects was
detected from the circumstellar envelope of the star (Loup et al. 1993).
Hrivnak \& Bieging (2005) noticed asymmetry and possible structure in the
CO J=2--1 and 4--3 lines. Recent imaging of I\,19475 with the Hubble Space
Telescope at optical wavelengths reveals a dusty quadrupolar nebula and
spherical halo, surrounding the central star (Sahai 2004; Sahai et al.
2006).

The rest of the paper is organized as follows. In \S\,2, we provide
observational
details of the ISO data, in \S\,3 we describe the ISO spectra, and the
re-construction of I\,19475's spectral energy distribution (SED) from these
as
well as broad-band photometric data, in \S\,4 we describe our modelling of
the
SED using a spherical dust radiative transfer code, in \S\,5 we discuss our
models, and in \S\,6, we present our conclusions.

\section{ISO Observations}

Infrared Space Observatory (ISO) observations of I\,19475 were
extracted from the ISO data archive. These include spectroscopic
observations 
made with the Short Wavelength Spectrometer (SWS) and the 
Long Wavelength Spectrometer (LWS) and spectrophotometric 
observations made with the imaging photopolarimeter (ISOPHOT) 
onboard ISO. An off-source measurement made with the LWS in order to 
estimate the background was also extracted from the archive. 
A log of the observations is given in Table 1. 

ISO SWS spectra have a wavelength coverage of 2.38$-$45.2 $\mu$m. Our 
spectra were obtained in the low resolution mode (AOT 01) of the SWS 
instrument (de Graauw et al. 1996) with a 33\arcsec $\times$ 20\arcsec
aperture.
Each SWS spectrum contains 12 subspectra, that each consist of two scans, 
one in the direction of decreasing wavelength (`up' scan) and the other in 
the direction of increasing wavelength (`down' scan). There are small
regions of 
overlap in wavelength between the sub-spectra. Each sub-spectrum is
recorded 
by 12 independent detectors.

LWS spectra extend from 43$-$197$\mu$m. The LWS observations were obtained
in LWS01 mode, covering the full spectral range at a resolution
($\lambda/\Delta\lambda$) of ~$\sim$ 200. The LWS circular field of view 
had a diameter of 84\arcsec. The ISO LWS instrument and its calibration 
are described in Clegg et al. (1996) and in Swinyard et al. (1996)
respectively. 

ISOPHOT observations were carried out using the spectrophotometer subsystem
PHT-S. PHT-S consists of two low-resolution 
grating spectrometers covering the wavelength ranges 2.5$-$4.9 $\mu$m
(PHT-SS) 
and 5.8$-$11.6 $\mu$m (PHT-SL)
and having a common entrance aperture of 24\arcsec $\times$ 24\arcsec. Each
channel
has a linear 64-element array of Si:Ga detectors. The spectral band width
(FWHM)
of a single detector is 0.0383 $\mu$m for PHT-SS and 0.0918 $\mu$m for
PHT-SL,
resulting in a mean $\lambda/\Delta\lambda$ of about 95 for both channels.
A
more detailed description of PHT-S has been given by Klaas et al. (1997).
In
the PHT40 observing mode spectrophotometry is performed simultaneously at
wavelengths 2.5$-$4.9 $\mu$m and 5.8$-$11.6 $\mu$m.
\clearpage
\begin{table}
\begin{center}
\caption{Log of ISO observations}
\begin{tabular}{|c|c|c|c|c|c|c|}
\hline
Object & Instrument & Date of Obs.& Duration of Obs. (s) & TDT$^{\rm a}$ &
Mode$^{b}$
& Speed$^{c}$ \\
\hline \hline
IRAS 19475+3119 & SWS & 12/11/1996 & 1140 & 36100905 & SWS01 & 1 \\
IRAS 19475+3119 & SWS & 19/04/1997 & 6538 & 52000931 & SWS01 & 4 \\
IRAS 19475+3119 & LWS & 12/11/1996 & 1266 & 36100904 & LWS01 & -- \\
Off-source      & LWS & 12/11/1996 & 1268 & 36100943 & LWS01 & -- \\
IRAS 19475+3119 & PHT-S & 19/04/1997 & 364 & 52000932 & PHT40 & -- \\
\hline
\end{tabular}

\noindent \parbox{14cm}{$^{\rm a}$TDT number uniquely identifies each ISO
observation. 
$^{\rm b}$Observing mode used. $^{\rm c}$Speed corresponds to the scan
speed of
observation.}
\end{center}
\end{table}
\clearpage
\section{Analysis}

Details of data reduction are described in Appendix A. The reduced SWS,
LWS, and
PHT-S spectra are shown in Figs.\ref{swsspec}, \ref{lwsspec} \&
\ref{phtsspec} ,
respectively. 

\subsection {Infrared spectral features}

\begin{figure}
\epsscale{1.15}
\plotone{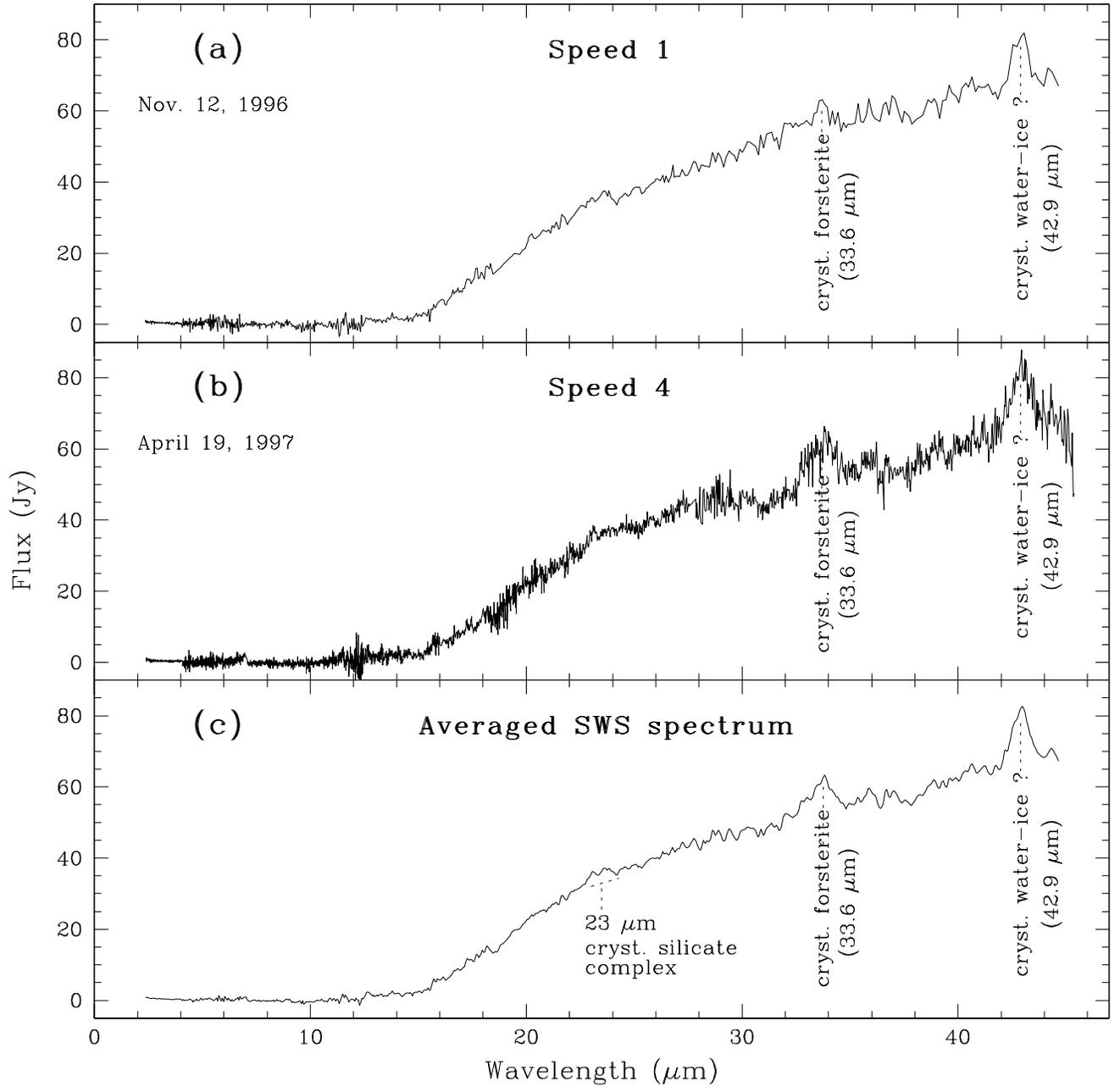}
\caption{SWS spectra of I\,19475 taken with (a) Speed 1 and (b) Speed 4
and (c) the averaged SWS spectrum of the star.}
\label{swsspec}
\end{figure}

Emission features at 33.6 $\mu$m and 43 $\mu$m were observed in the SWS
spectra
(Fig.\,\ref{swsspec}). The 33.6 $\mu$m feature is attributed to forsterite,
a 
crystalline silicate (Waters et al. 1996; Waters \& Molster 1999; Gauba \&
Parthasarathy 2004). Both pyroxenes (such as enstatite) and water-ice 
show features at $\sim$ 43 $\mu$m. ISO observations of post-AGB stars have
shown that
the 43 $\mu$m feature is almost always accompanied with a prominent
emission feature
at 40.5 $\mu$m attributed to pyroxenes such as enstatite and/or diopside
(Molster et al. 2002b). The non-detection of the 40.5 $\mu$m feature in
I\,19475 is
therefore puzzling, suggesting that pyroxenes are not (or only very weakly)
present
in this object

The only other post-AGB object which shows a spectrum in the $\sim$40
$\mu$m region
similar to that of I\,19475, is HD161796 (Molster et al. 2002b), which,
like 
I\,19475, is a high latitude F3Ib post-AGB star with a detached
circumstellar dust
envelope (Parthasarathy \& Pottasch 1986; Hoogzaad et al. 2002).  HD161796
shows a 
prominent 43 $\mu$m feature and a weak 40.5 $\mu$m feature -- in order to
fit these 
features, Molster et al. (2002c) need a combination of pyroxenes and
crystalline
water ice, with the latter dominating the $\sim$40 $\mu$m emission complex.
This is
because in the $\sim$40 $\mu$m region, crystalline water ice contributes a
feature
only at 43 $\mu$m, hence its dominant presence helps to reduce the
intensity of the
40.5 $\mu$m feature (produced by pyroxene only) relative to the 43 $\mu$m
one
(produced by
pyroxene and ice). The presence of crystalline water ice in HD161796 is supported by
the
prominent 62 $\mu$m feature in its LWS spectrum.

Similarly, if the 43 $\mu$m feature in I\,19475 is due to (or dominated by)
crystalline water ice, then one would expect to see a crystalline water-ice feature
at $\sim$ 62 $\mu$m. We now examine the possibility that the latter is not seen
because it may be much weaker compared to the 43 $\mu$m feature if the
ice mantles are deposited onto the dust grains at a temperature above 110\,K (Smith
et al. 1994), by as much as a factor of 3.1. And in fact, in most PPNs, the
continuum-subtracted peak intensity of the 62 $\mu$m crystalline water-ice feature
when present, is about a factor of $\sim$3 lower than the 43 $\mu$m feature (see
e.g. Molster et al. 2002c). The continuum-subtracted peak intensity of the 43 $\mu$m
feature in our spectra is 14.4 $\pm$ 1.4 Jy (SWS 36100905) and 17.6 $\pm$ 0.8 Jy
(SWS 52000931) respectively. Hence we would expect the $\sim$ 62 $\mu$m feature in
I\,19475 to have a peak intensity of about 4.6 Jy; given that the expected width of
this feature, both from laboratory and astronomical spectra, is quite large (about
8 $\mu$m) this feature should have been detected in our spectra. We conclude that
the 62 $\mu$m feature is probably not present in I\,19475; the identification of the
43 $\mu$m emission feature in our spectra as due to crystalline water ice is thus 
inconclusive.

Unlike crystalline water ice, amorphous water ice produces a single broad feature
with a peak at $\sim$ 46 $\mu$m (Smith et al. 1994). Molster et al. (2002c) suggest
that some amorphous water ice may be present in HD161796 in order to explain modest
discrepancies between their model and the data in the $\sim$40 $\mu$m wavelength
region. However, in I\,19475, the 43 $\mu$m feature is relatively narrow, with a
peak at 42.9 $\pm$ 0.04 $\mu$m. The sharpness of this feature suggests that it is
unlikely that amorphous water ice has a substantial presence in I\,19475. However,
without detailed modelling of this feature a small contribution from amorphous water
ice cannot be ruled out. This is because even if we assume that all ice condenses in
crystalline form (requiring extreme conditions in the AGB outflow), UV radiation
from the interstellar medium (ISM) can begin to amorphisize the ice (Dijkstra et al.
2003b and references therein).

\begin{figure}
\epsscale{1.15}
\plotone{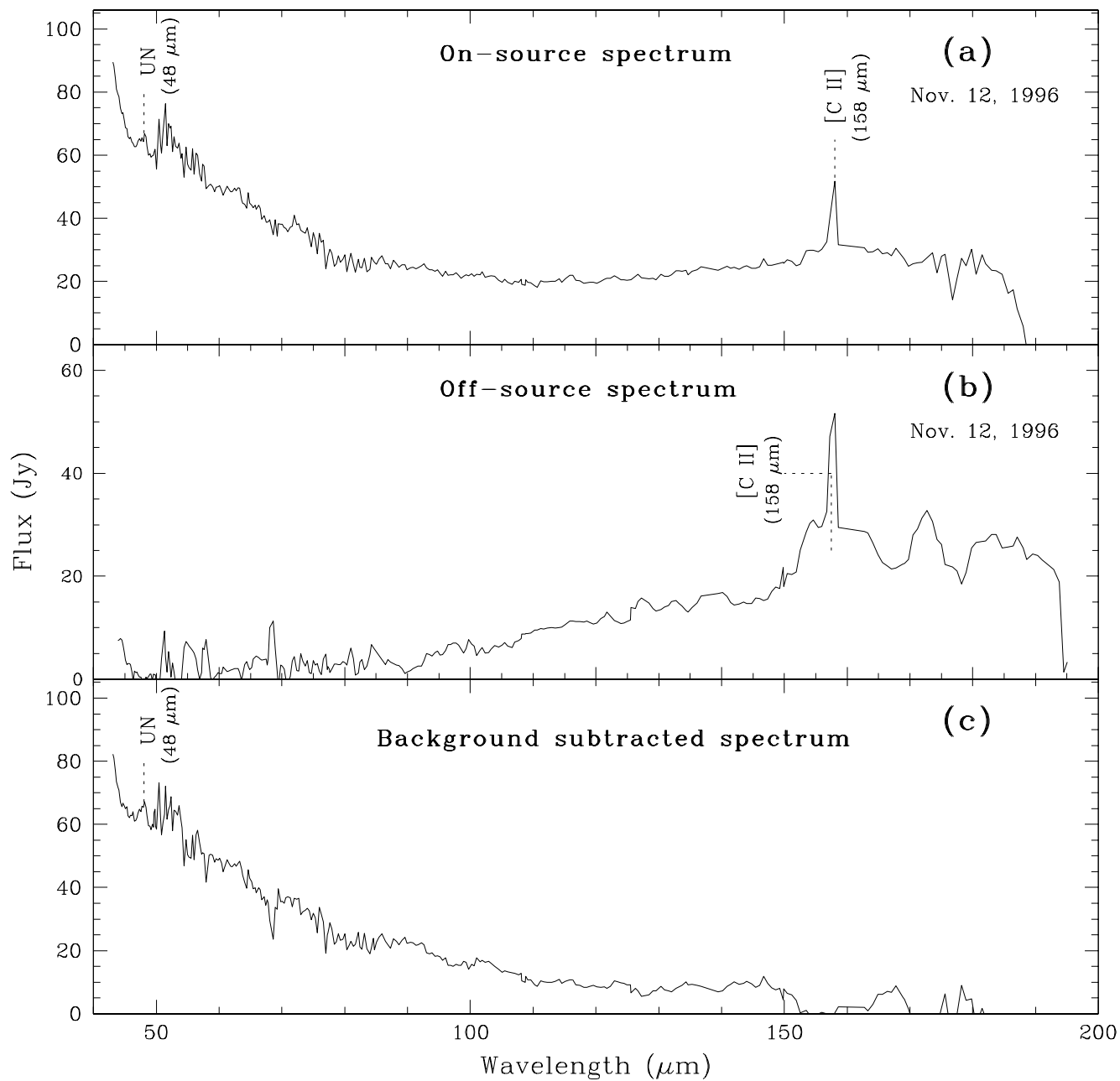}
\caption{(a) The observed LWS spectrum of I\,19475, (b) off-source
spectrum, and (c) source spectrum after subtraction of the
off-source data. The 48 $\mu$m emission feature seen in the LWS spectrum is
as yet 
unidentified (UN)}
\label{lwsspec}
\end{figure}

\begin{figure}
\includegraphics[scale=0.4,angle=-90]{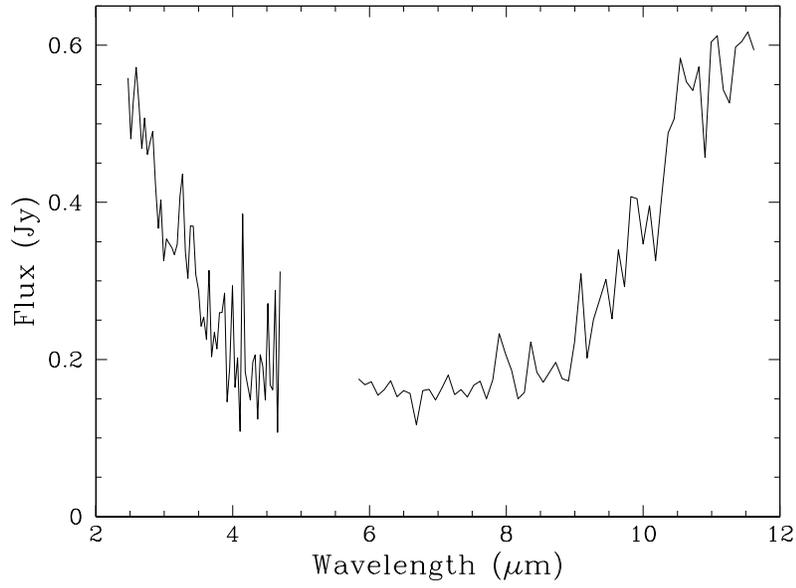}
\caption{The PHT-S spectrum of I\,19475 recorded with the low resolution
grating 
spectrometers PHT-SS (2.5 $-$ 4.9 $\mu$m) and PHT-SL (5.8 $-$ 11.6 $\mu$m)
onboard
ISO.}
\label{phtsspec}
\end{figure}

In the averaged SWS spectrum (Fig.\,\ref{swsspec}c), a weak emission feature at
$\sim$ 23 $\mu$m attributed to crystalline silicates (Molster et al. 2002b) becomes
obvious. An emission feature at 48 $\mu$m was detected in the LWS spectrum of
I\,19475 (Fig.\,\ref{lwsspec}). The 48 $\mu$m feature has been observed in several
stars (see e.g. Molster et al., 2002b). Ferrarotti et al. (2000) identified
this feature with FeSi. However, this identification remains unconvincing since the
formation of FeSi would require an environment where C/O $\sim$ 1, whereas the 48
$\mu$m feature has been observed in several O-rich stars as well (Molster et al.
2002b). Furthermore, it is as yet not clear if this spectral
feature is an artifact in the LWS instrument spectral response function, or a real
feature (T. Lim, private communication 2006). 

The presence of [CII] emission at 158 $\mu$m in both the on-source and
off-source
spectra, combined with the absence of
[OI] at 63 $\mu$m in the on-source LWS spectrum (Fig.\,\ref{lwsspec})
suggests
that 
the [CII] emission can be attributed entirely to contamination from 
the ISM (see e.g. Castro-Carrizo et al. 2001).
\clearpage

\clearpage
\subsection {Spectral energy distribution (SED)}

\begin{figure}
\includegraphics[scale=0.7,angle=90]{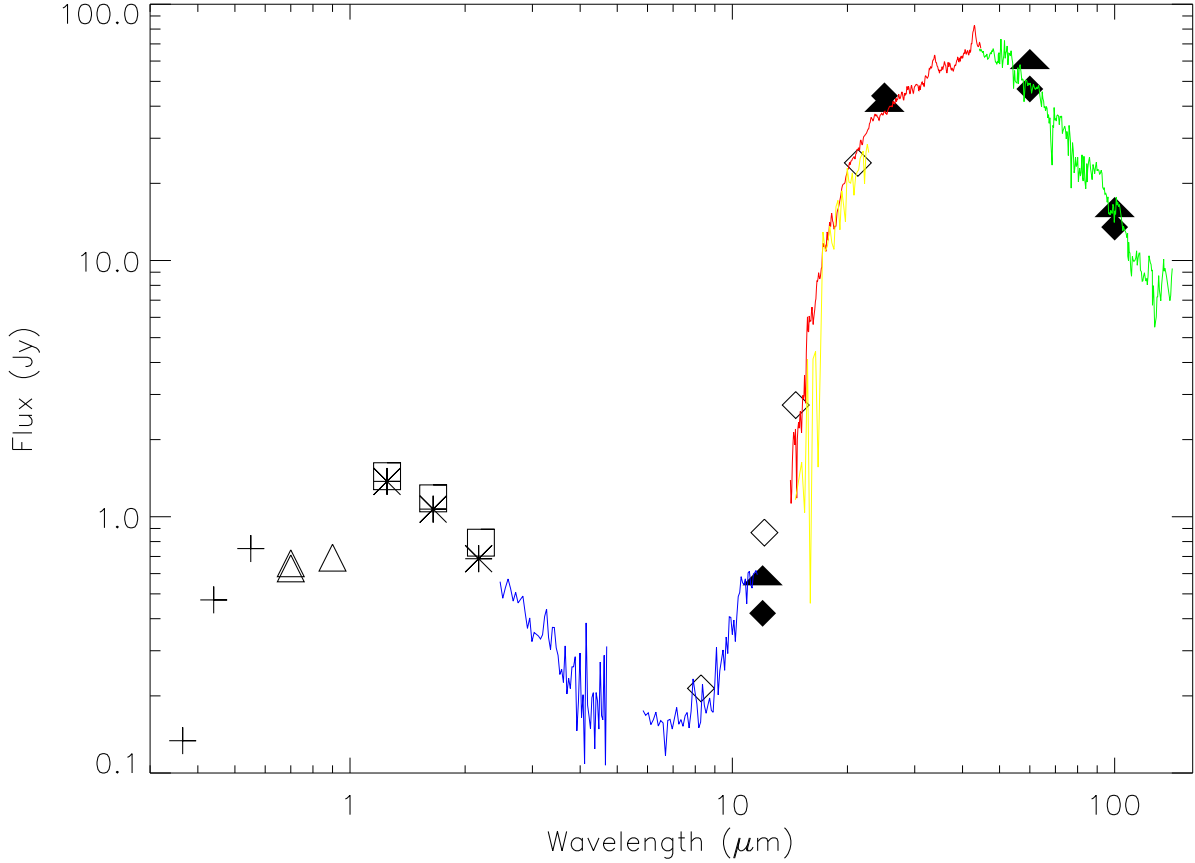}
\caption{The observed SED of I\,19475. The plotted data are not corrected
for interstellar and circumstellar reddening. U, B, V (plus symbols), R, I
(triangles) 
fluxes are plotted along with J, H, K fluxes (open squares: data from 
Garc\'ia-Lario et al., 1997; asterisk: 2MASS data), MSX (diamonds) 
and IRAS data (filled triangles : IRAS fluxes from the Point Source
Catalog;
filled diamonds : color-corrected IRAS fluxes). Colors highlight the PHT-S
(blue),
IRAS LRS (yellow), SWS (red) 
and LWS (green) data.}
\label{colorspec}
\end{figure}

Both SWS spectra of I\,19475 show similar flux 
distributions (Figs.\ref{swsspec}a and b). Therefore, for further analysis,
the
two SWS observations were averaged together after rebinning the 
Speed 4 data to the resolution of the Speed 1 SWS data (Fig. 1c).
The averaged SWS data were combined with the PHT-S data
(Fig.\,\ref{phtsspec}) and
the 
background subtracted LWS spectrum of the star (Fig.\,\ref{lwsspec}c).
The SWS and LWS data showed good agreement in the overlap region. The SWS
data
below 14 $\mu$m and the LWS data beyond 143 $\mu$m are noisy (see Appendix
A and
Fig.\,\ref{lwsspec}c) and were excluded from further analysis. To
reconstruct the
SED, 
the ISO data were combined with
the available U,B,V (Reed 2001), R,I (Monet et al. 2003), 
J,H,K (2Micron All Sky Survey (2MASS); Garc\'ia-Lario et al. 1997)
magnitudes, 
MSX (Midcourse Space Experiment) fluxes and IRAS (Infrared Astronomical
Satellite) 
fluxes of the star (Tables 2, 3 and 4).  
Color-corrected IRAS fluxes (Table 4) were estimated using the
IRAS Explanatory Supplement. We also retrieved the IRAS low resolution
spectrum
(LRS) from the University of Calgary database 
(http://www.iras.ucalgary.ca/$\sim$volk/getlrs\_plot.html) using the
``corrected
raw text" 
option which applies the absolute calibration corrections from Cohen et al.
(1992) 
to the raw IRAS LRS data. The short-wavelength (7.7$-$13.4 $\mu$m) section
of
the LRS spectrum is very noisy, and has not been included. 
The full optical--to--far-infrared SED of the star is
shown in Fig.\,\ref{colorspec}. 
\clearpage
\clearpage

\begin{table}
\begin{center}
\caption{Photometric data of I\,19475}
\begin{tabular}{|c|c|c|c|c|c|c|c|}
\hline
U & B & V & R & I & J & H & K \\
mag & mag & mag & mag & mag & mag & mag & mag \\
\hline
10.404 & 9.97 & 9.312 & 9.14$^{\rm a}$ & 8.78 & 7.73$^{\rm c}$ & 7.41$^{\rm
c}$ &
7.25$^{\rm c}$\\ 
       &      &       & 9.10$^{\rm b}$ &      & 7.773$^{\rm d}$ &
7.493$^{\rm d}$
&
7.366$^{\rm d}$\\ 
\hline
\end{tabular}

\noindent \parbox{14cm}{a and b are R-band observations at 
epochs 1 and 2 respectively (Monet et al., 2003).  
$^{\rm c}$:Garc\'ia-Lario et al. (1997); $^{\rm d}$: 2MASS data}
\end{center}
\end{table}

\begin{table}
\begin{center}
\caption{MSX data}
\begin{tabular}{|c|c|c|c|}
\hline
\multicolumn{4}{|c|}{MSX fluxes in Janskys} \\ \cline{1-4}
Band A & Band C & Band D & Band E \\
8.28 $\mu$m & 12.13 $\mu$m & 14.65 $\mu$m & 21.34 $\mu$m \\ 
\hline
 0.214 & 0.867 & 2.728 & 24.041 \\
\hline
\end{tabular}
\end{center}
\end{table}

\begin{table}[t]
\begin{center}
\caption{IRAS data}
\begin{tabular}{|c|c|c|c|}
\hline
\multicolumn{4}{|c|}{IRAS fluxes in Janskys} \\ \cline{1-4}
12 $\mu$m & 25 $\mu$m & 60 $\mu$m & 100 $\mu$m \\
\hline
0.54 & 37.99 & 55.83 & 14.76 \\
0.42$^{\dagger}$ & 43.9$^{\dagger}$ & 46.7$^{\dagger}$ & 13.5$^{\dagger}$
\\
\hline
\end{tabular}

\noindent \parbox{14cm}{$^{\dagger}$: Color-corrected IRAS fluxes}
\end{center}
\end{table}
\clearpage
\section {Dust Shell Modelling of the SED}

We now describe detailed modelling of the SED of I\,19475, using the 1-D
radiative
transfer code, DUSTY (Ivezi\'c et al. 1999). DUSTY assumes a spherical
geometry
and supports various analytical forms for the dust density distribution.
The
input parameters required for the DUSTY code are the dust temperature on
the
inner shell boundary (T$_{\rm d}$), the relative shell thickness (Y = ratio
of
the outer to the inner shell radius), the optical depth ($\tau$) at a
specified
wavelength, the dust grain composition and grain size distribution, and the
spectrum of the heating radiation from the central source.

\subsection {Distance estimates}

Although the distance to the source is not a required input for DUSTY, we
need it
in order to determine the physical size and total mass of the dust shell.
Various distance estimates for I\,19475 are available in
literature. Likkel et al. (1987) used a kinematic distance 
estimate of 6 kpc. Omont et al. (1993) adopted a luminosity of
10$^{4}$ L$_{\odot}$ and estimated a distance of 4.8 kpc. 
Hrivnak \& Bieging (2005) estimated 
a distance of 4.9 kpc to the object, assuming a luminosity of 8300
L$_\odot$,
appropriate for a core mass 
of 0.63 M$_\odot$ (Blocker 1995) -- we have adopted their values for the
models
discussed in this paper. 

\subsection {Reddening}

The total extinction towards the star, 
(E(B$-$V)=0.41), was estimated from the 
difference between the observed B$-$V (= 0.66) value and intrinsic 
B$-$V ((B$-$V)$_{\rm O}$) value for the optical spectral type 
(F3Ia) of the star. For (B$-$V)$_{\rm O}$ we used a value of
$+$0.25 after interpolating between (B$-$V)$_{\rm O}$ for a
F2Ia and a F5Ia star (Schmidt-Kaler 1982). Thus, A$_{\rm v}$=1.3, using 
R$_{\rm v}$ = 3.1, where, R$_{\rm v}$ = A$_{\rm v}$/E(B$-$V) (Rieke \&
Lebofsky
1985). Alternatively, using the numerical algorithm provided by Hakkila et
al.
(1997), which computes
the 3-dimensional visual interstellar extinction and its error from inputs
of
Galactic longitude and latitude, and distance, from a synthesis of several
published studies, we find an interstellar extinction of $A_V=1.5\pm0.5$.
However,
at the low
galactic latitude (b = +2.73) of the star, galactic extinction maps
(e.g. Schlegel et al. 1998) do not provide very reliable estimates
of the interstellar extinction. We have corrected 
the observed optical and near-infrared magnitudes for an extinction A$_{\rm
v}$=1.3, using the standard extinction 
laws by Rieke \& Lebofsky (1985).

\subsection {Stellar model atmosphere parameters}

A Kurucz model atmosphere was used as the input stellar 
spectrum. 
Several estimates for the effective temperature (T$_{\rm eff}$)
and gravity (log g) of I\,19475 are found in literature:
T$_{\rm eff}$ = 7200 K, log g = 0.5 (Klochkova et al. 2002),
T$_{\rm eff}$ = 7750 K, log g = 1.0 (Arellano Ferro et al. 2001) and
T$_{\rm eff}$ = 7500 K, log g = 0.5 (Sivarani et al. 2001). In
this paper, for the purpose of modelling the SED, 
we adopted a Kurucz model atmosphere with T$_{\rm eff}$ = 7500 K and 
log g = 1.0.

\subsection {Model Fits}

In the sections below we discuss the
various model fits to the observed data. The input and output parameters
corresponding to these models are listed in Table 5.

The modelled flux density (reddened due to circumstellar dust) is scaled to
match 
the observed flux density in the K-band (2.2 $\mu$m) (Garci\'a-Lario et al.
1997), corrected for
the total (interstellar and circumstellar) extinction. 
Garci\'a-Lario et al.'s measurement is consistent, within observational
uncertainties, with the 2MASS K-band photometry which is fainter by 0.11
magnitudes. The total extinction at the K-band was estimated to be 0.14
magnitudes.
For our best fit DUSTY model (discussed later), the circumstellar
extinction at the K-band (in optical depth units) is 0.04.

We have chosen to use the 2.2 $\mu$m flux rather than, e.g., an optical
flux, for
scaling the model to the data, in order to minimise the sensitivity of our
results
to the following two modelling uncertainties. First, we do not have a very
reliable estimate for the interstellar extinction, and second, we are using
a 1-D
model for an object where a significant part of the scattered light
contribution
at short wavelengths comes from non-spherical components. 
At 2.2 $\mu$m, the wavelength is (i) sufficiently long that the scattered
light
contribution is small compared to the direct (but extincted) stellar
contribution,
and the (uncertain)
extinction correction is small, and is (ii) sufficiently short, that the
dust
shell does not contribute significantly via emission (so our scaling is not
affected by the dust properties which are to be derived from the model
fitting).

\subsubsection {Grain types and size distribution}

\begin{figure}
\includegraphics[scale=0.93]{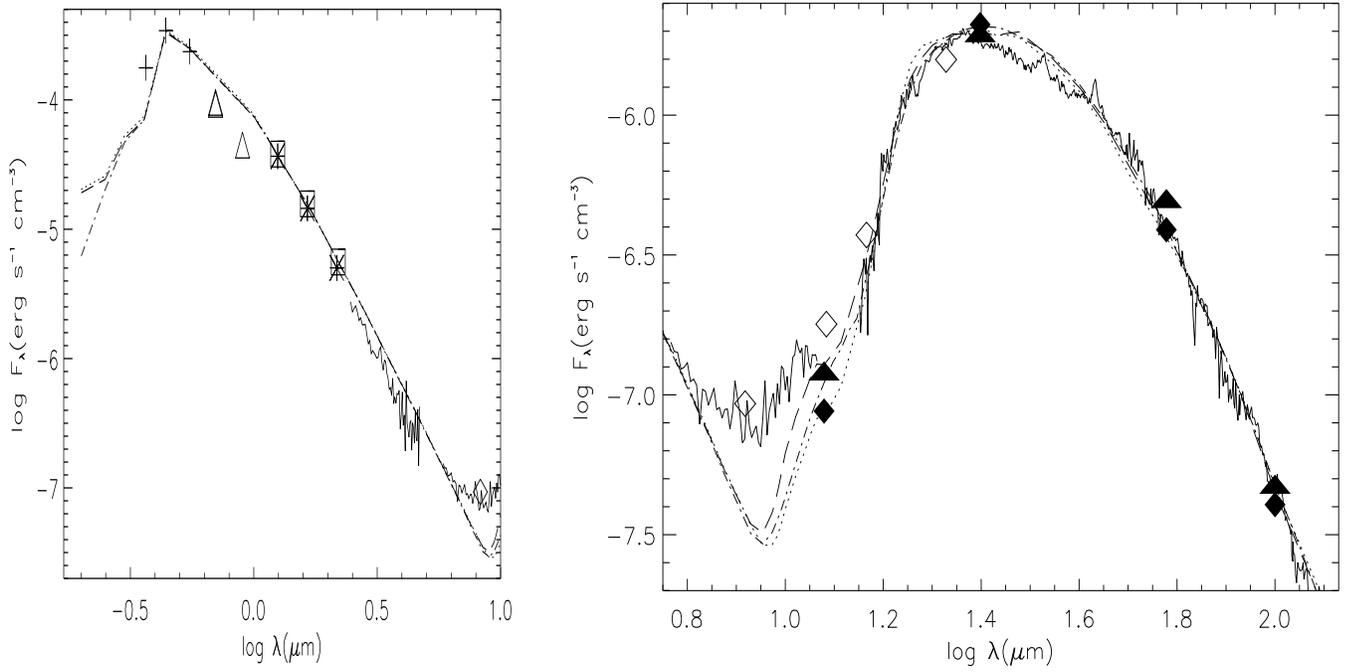}
\caption{Model fits to the observed SED using Sil-Ow (dashed curve), 
Sil-Oc (dotted curve) and Sil-DL (dashed-dotted curve) grain types
(model numbers 1, 2 and 3 in Table 5). The meanings of 
the symbols are as in Fig.\,\ref{colorspec}. The PHT-S, SWS
and LWS spectra are shown as solid curves. 
}
\label{mod-graintypes}
\end{figure}

We have selected grain types for our models from the six different grain
types
whose
constants are directly available in DUSTY: ``warm'' (Sil-Ow) and ``cold''
(Sil-Oc) silicates from Ossenkopf et al. (1992), silicates and 
graphites (Sil-DL and grf-DL) from Draine and Lee (1984), 
amorphous carbon (amC-Hn) from Hanner (1988) and SiC (SiC-Pg) from
P\'egouri\'e) (1988). We find that using silicate grains  
(Sil-Ow, Sil-Oc, Sil-DL) provide good fits to the SED. The use of 
silicate grains is also supported by our discovery of silicate features
(crystalline 
forsterite and pyroxenes) in the SWS spectrum of I\,19475
(Fig.\,\ref{swsspec}).
Fig.\,\ref{mod-graintypes} shows the modelled fits to the observed SED
using
Sil-Ow,
Sil-Oc and
Sil-DL grains.  Optical constants of the Sil-DL grains 
from Draine and Lee (1984) are based on a combination of
laboratory measurements and astronomical observations. 
Optical constants of the warm oxygen-deficient 
(Sil-Ow) and cool oxygen-rich (Sil-Oc) silicates are based on
observational determinations of the opacities of circumstellar and 
interstellar silicates as well as laboratory data (Ossenkopf et al. 1992). 
In estimating the optical constants of Sil-Ow and Sil-Oc, the effects of 
small spherical inclusions of various materials (oxides, sulfides,
carbides,
amorphous 
carbon and metallic iron) upon silicate opacities were taken into account. 
Fig.\,\ref{mod-graintypes} shows that using Sil-Oc has the effect of
producing
slightly higher 
flux at $\sim$ 18 $-$ 25$\mu$m. Sil-Ow grains appear to provide the best
possible
fit to the SED from the near to the far-infrared. We therefore adopted 
Sil-Ow for the purpose of modelling.

The standard Mathis, Rumpl, Nordsieck (MRN) (Mathis et al. 1977) power law 
was assumed for the grain size distribution, 
i.e., n(a) $\propto$ a$^{\rm -q}$ for a$_{\rm min}\le$a$_{\rm max}$
with q = 3.5, a$_{\rm min}$ = 0.005 $\mu$m and a$_{\rm max}$ = 0.25 $\mu$m.
\clearpage
\clearpage
\subsubsection {Density distribution}

\begin{figure}
\includegraphics[scale=0.93]{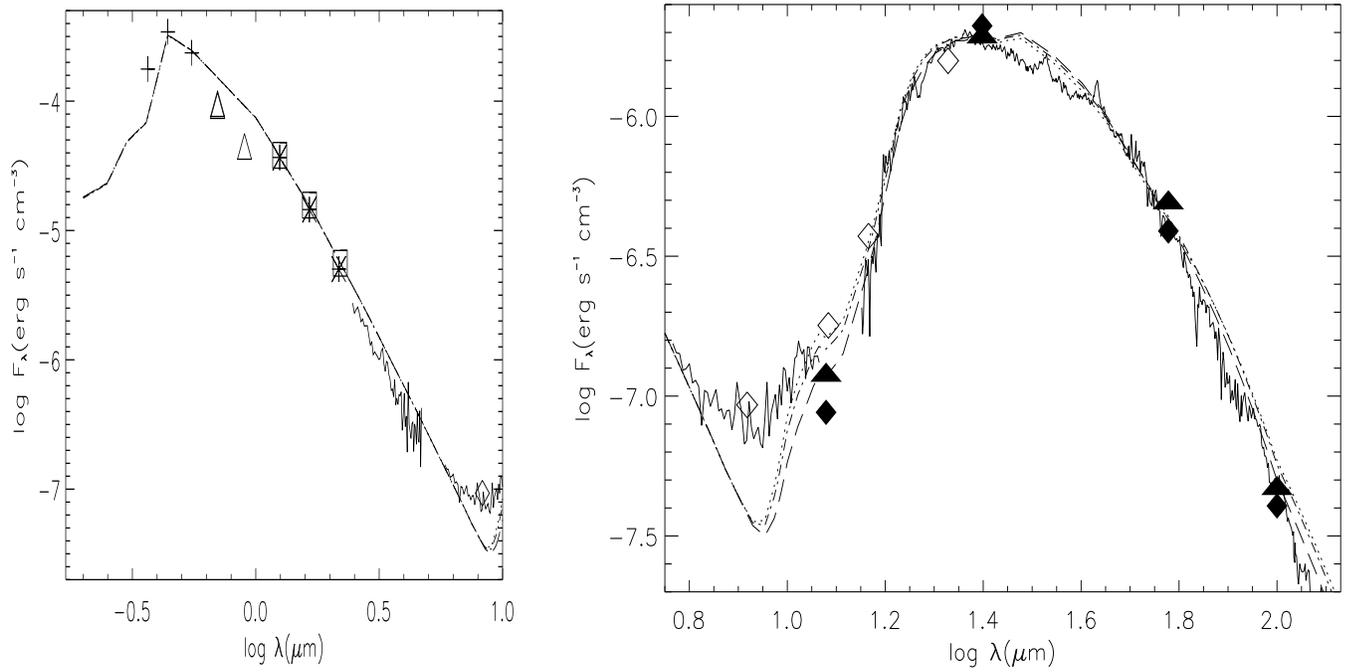}
\caption{As in Fig.\,\ref{mod-graintypes}, but showing model fits using a
r$^{-2}$
density distribution
and relative
shell thickness of 5 (dashed line), 10 (dashed-dotted) and 15 (dotted)
(model
numbers
4, 5 and 6 in Table 5).}
\label{mod-power2}
\end{figure}

\begin{figure}
\includegraphics[scale=0.93]{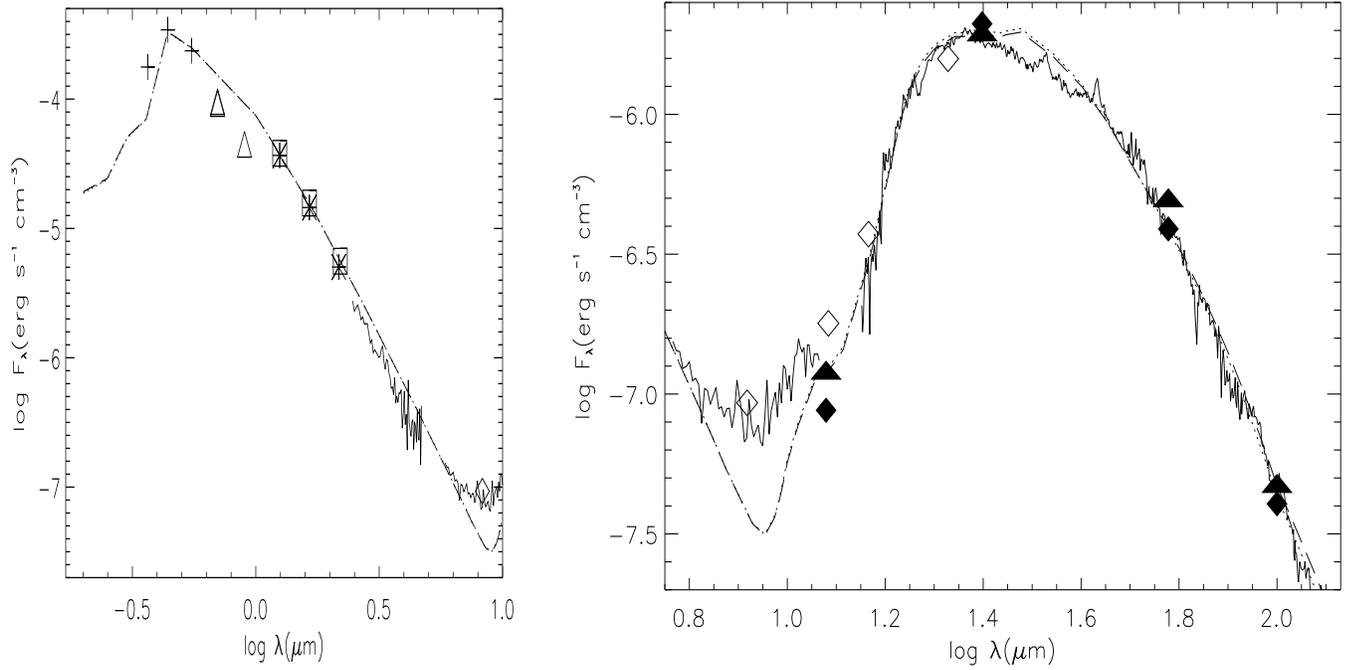}
\caption{As in Fig.\,\ref{mod-graintypes}, but showing model fits using a
r$^{-3}$
density distribution
and relative 
shell thickness of 5 (dotted) and 15 (dashed) (model numbers
7 and 8 in Table 5).}
\label{mod-power3}
\end{figure}

We tried two power laws, r$^{-2}$ and r$^{-3}$, for the 
density distribution. The best fits in the case of the r$^{-2}$ 
power law for Y of 5, 10 and 15 are shown
in Fig.\,\ref{mod-power2}. Using the r$^{-2}$ power law, we find that even 
with Y as small as 5, the modelled flux systematically exceeds the LWS
flux beyond $\sim$70 $\mu$m. Adopting an r$^{-3}$ density distribution and 
Y = 5, produces a good fit to the SED upto the limit 
of the LWS spectrum. For Y $>$ 5, the modelled flux 
again begins to exceed the observed LWS flux beyond
100 $\mu$m. The best fit models
using r$^{-3}$ power law and Y = 5 and 15 are shown in
Fig.\,\ref{mod-power3}. 
\clearpage

\clearpage
\subsubsection {Additional constraints on the models}

Recent HST images of I\,19475 (Sahai et al. 2006) show a r$^{-3}$ halo with an inner
radius of $\sim$ 1.5\arcsec; the surface-brightness limited value of the outer
radius is $\sim$ 5\arcsec. For our assumed distance (4.9 kpc), these values
correspond to R$_{\rm in}$ = 1.1 $\times$ 10$^{17}$cm and R$_{\rm out}$ = 3.6
$\times$ 10$^{17}$cm, which are in good agreement with our model values -- e.g., in
Model 7, R$_{\rm in}$ = 8.8 $\times$ 10$^{16}$cm and R$_{\rm out}$ = 4.4 $\times$
10$^{17}$cm.

Additional constraints on our model are provided by the observed flux of I\,19475 at
sub-mm wavelengths. Gledhill et al. (2002) measured a flux density of 34$\pm$5.4 mJy
at 850 $\mu$m using a beamwidth of 13\arcsec. The wide-band filter used by Gledhill
et al. (2002) includes the CO J=3--2 line in its range. However, we find, using
formulae from Seaquist et al. (2004) for calculating the line contribution to the
filter, and an estimate of the CO J=3--2 line flux from a multi-transition model fit
to CO data on I\,19475, that the line emission is not likely to contribute more than
20-30\% of the observed flux (Sahai et al. 2006). Hence the sub-mm flux of I\,19475
is $\gtrsim$20 mJy. For our best-fit model (Model 7 with r$^{-3}$ power law and Y=5)
we integrated the model surface brightness distribution (in Jy arcsec$^{-2}$) at 850
$\mu$m output by DUSTY, using a 13\arcsec~diameter aperture to obtain the 850 $\mu$m
model flux density. DUSTY outputs the surface brightness distribution as a function
of normalized (in units of R$_{\rm in}$) radial distance from the central source.
The radial distances were converted into angular radii ($\theta$) using our adopted
distance of 4.9 kpc to the source. Thus Y=5 corresponds to a shell size of
12\arcsec, for which we find a 850 $\mu$m flux density of 10 mJy, significantly
lower than the observed value. We discuss the implications of this discrepancy for
our modelling results in \S\,5.

\clearpage
\begin{figure}
\includegraphics[scale=0.93]{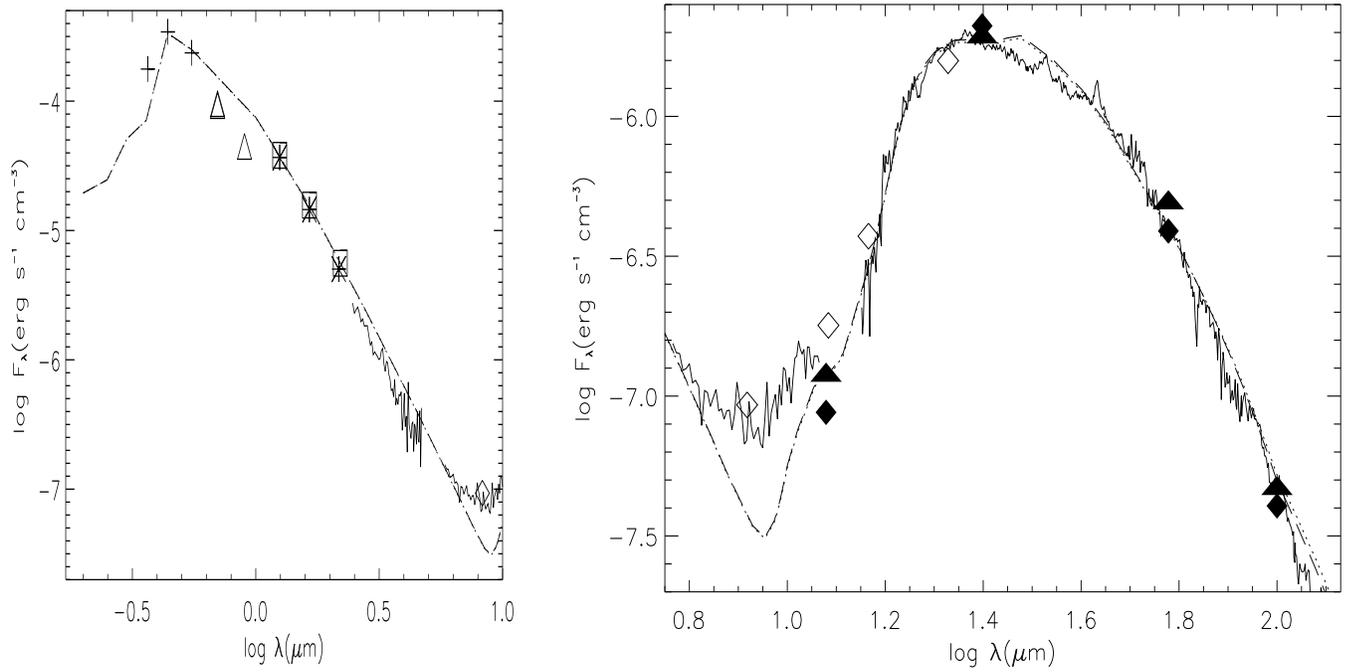}
\caption{As in Fig.\,\ref{mod-graintypes}, but showing model fits with an inner shell
of relative thickness 5, and an outer shell of relative thickness 15 (dashed) and 30
(dotted) (model numbers 9 and 10 in Table 5).}
\label{mod-outshell}
\end{figure}
\clearpage
\begin{table}
\begin{center}
\caption{Input and output parameters for DUSTY model fits to the observed
data}
\begin{tabular}{|c|c|c|c|c|c|c|c|c|}
\hline
Model & Grain & Density & T$_{\rm d}$ & \multicolumn{2}{c|}{$\tau$} &
R$_{\rm
in}$$^{\dagger}$ &
Shell     & Shell Mass\\
number& type  & law     &   (K)       & at 0.55 $\mu$m & at 100 $\mu$m &
(cm)                    
&
thickness & (M$_{\odot}$)\\ 
\hline \hline

1  & Sil-Ow & r$^{-3}$ &  95 & 1.00 & 9.874 $\times$ 10$^{-4}$ & 8.54
$\times
10^{16}$ & 10 & 1.24 \\
2  & Sil-Oc & r$^{-3}$ &  88 & 1.00 & 8.404 $\times$ 10$^{-4}$ & 1.00
$\times
10^{17}$ & 10 & 1.45 \\ 
3  & Sil-DL & r$^{-3}$ &  93 & 1.50 & 2.631 $\times$ 10$^{-3}$ & 5.70
$\times
10^{16}$ & 10 & 1.47 \\
4  & Sil-Ow & r$^{-2}$ &  96 & 1.05 & 1.037 $\times$ 10$^{-3}$ & 8.06
$\times
10^{16}$ &  5 & 1.25 \\
5  & Sil-Ow & r$^{-2}$ & 100 & 1.10 & 1.086 $\times$ 10$^{-3}$ & 7.12
$\times
10^{16}$ & 10 & 2.05 \\ 
6  & Sil-Ow & r$^{-2}$ & 102 & 1.10 & 1.086 $\times$ 10$^{-3}$ & 6.71
$\times
10^{16}$ & 15 & 2.72 \\
7  & Sil-Ow & r$^{-3}$ &  94 & 1.00 & 9.874 $\times$ 10$^{-4}$ & 8.84
$\times
10^{16}$ &  5 & 0.96 \\
8  & Sil-Ow & r$^{-3}$ &  94 & 1.00 & 9.874 $\times$ 10$^{-4}$ & 8.80
$\times
10^{16}$ & 15 & 1.54 \\
9  & Sil-Ow & r$^{-3}$, r$^{-2}$ & 94 & 1.00 & 9.874 $\times$ 10$^{-4}$ &
8.77
$\times 10^{16}$ & 5, 15 &
2.01 \\
10 & Sil-Ow & r$^{-3}$, r$^{-2}$ & 94 & 1.00 & 9.874 $\times$ 10$^{-4}$ &
8.76
$\times 10^{16}$ & 5, 30 &
3.63 \\ 
\hline
\end{tabular}
\clearpage
\noindent \parbox{14cm}{$^{\dagger}$: The inner radius of the dust shell
where the dust temperature (T$_{\rm d}$) is specified. R$_{\rm in}$ as
output 
by DUSTY scales as L$^{1/2}$ where, L is the luminosity. The R$_{\rm in}$
values 
listed here are for a central star luminosity of 8300 L$_{\odot}$.}
\end{center}
\end{table}

\subsubsection{Dust shell mass}

The mass in the circumstellar dust shell (M$_{\rm s}$) was computed for each of the
models (Table 5). For an r$^{-3}$ power law dust density law,
\begin{eqnarray}
M_{\rm s} = 8 \pi R_{\rm in}^{2} (1-Y^{-2})^{-1} ln Y
(\tau_{100}/\kappa_{100})\,\delta,
\label{r3}
\end{eqnarray}
where, R$_{\rm in}$ is the inner radius of the dust shell (Table 5), Y is the
relative shell thickness (R$_{\rm out}$/R$_{\rm in}$), $\tau_{100}$ is the shell
optical depth at 100 $\mu$m, $\kappa_{100}$ is the dust mass absorption coefficient
at 100 $\mu$m, and $\delta$ is the gas-to-dust ratio. When the density distribution
follows a r$^{-2}$ power law,
\begin{eqnarray} 
M_{\rm s} = 4 \pi R_{\rm in}^{2} Y (\tau_{100}/\kappa_{100})\,\delta.
\end{eqnarray} 
For the models with an r$^{-3}$ inner shell, and an r$^{-2}$ outer shell (see \S\,5),
\begin{eqnarray}
M_{\rm s} = 4 \pi R_{\rm in}^{2} (\tau_{100}/\kappa_{100})\,\delta[ln Y1 +
(Y2/Y1-1)]/g(Y1,Y2),
\end{eqnarray}
where g(Y1,Y2)=(1-1/Y1$^2$)/2 + (1-Y1/Y2)/Y1$^2$, and Y1 and Y2 are the relative
shell thicknesses of the inner and outer shell, respectively. The value of the dust
mass absorption coefficient $\kappa$ at far-infrared wavelength is poorly known
(e.g. Jura 1986). We use $\kappa_{100}$ = 34 cm$^{2}$ gm$^{-1}$, appropriate for the
silicate dust used in our models (e.g. tabulated values by Ossenkopf at
http://hera.ph1.uni-koeln.de/$\sim$ossk/Jena/tables.html), and $\delta$=200 (a
typical value for the ejecta from AGB stars) to derive a shell mass of 1\,\ms~for
our best-fit model (shell masses for this model and others are given in Table 5).
The corresponding mass-loss rate is 3.4$\times 10^{-4}$\my~(6.7$\times 10^{-5}$\my)
at the inner (outer) radius of the shell, assuming a constant expansion velocity of
16\,\kms~based on CO line observations (Sahai et al. 2006).

Our derived shell mass should really be considered a lower limit because
the
model SED is not very sensitive to cooler dust at large radii. This is
demonstrated by the fact that the quality of the fits to the SED from the
additional models with much larger outer radii compared to Model 7 (listed
in
Table 5), although poorer than the latter, is not drastically worse.

\section{Discussion}

Our model fits to the SED of I\,19475 covering the optical to the far-infrared
wavelength range are reasonably satisfactory. The radial density distribution and
physical size of the dust shell are also in reasonable agreement with the imaging
data from HST. Since modelling of CO J=4-3 and J=2-1 line emission from I\,19475
also suggests a r$^{-3}$ radial-density shell (Hrivnak \& Bieging 2005), the
r$^{-3}$ density law for I\,19475's dusty, molecular envelope has strong support.
The exponent of the density law characterising the ambient circumstellar medium has
important implications for the primary process which results in the observed shapes
of PPNs and PNs (Sahai \& Trauger 1998) -- namely, the interaction of fast
collimated post-AGB wind with this medium. For example, the forward shock produced
by a fast wind expanding into an r$^{-2}$ medium moves at constant velocity, but in
an r$^{-3}$ medium, the shock accelerates and is thus subject to Rayleigh-Taylor
instabilities. Thus, the dynamical age and the morphology of the lobes in PPNs and
PNs are expected to be closely related to the steepness of the radial density
power-law distribution.

We note however that there is a caveat to the agreement in the density law exponent
between our dust model and the CO models -- the latter give a mass-loss rate of
5.2$\times 10^{-5}$\my at an inner radius R$_{\rm in}=3.9\times 10^{16}$cm based on
(unpublished) mid-infrared imaging data, which is significantly lower than our
estimate at an even larger value of R$_{\rm in}$. Since the CO models for inferring
the mass-loss rate depend on a number of poorly-known quantities such as the radial
kinetic temperature distribution, the inner and outer radius of the CO envelope, and
the fractional CO abundance, we believe that our modelling of the dust emission
provide an independent and probably more robust estimate of the shell mass and
mass-loss rate.

The discrepancy between the observed and model sub-mm flux is intriguing, suggesting
that there are either (1) significant quantities of cooler grains in I\,19475 than
included in our model, and/or (2) the dust emissivity at sub-millimeter wavelengths
is significantly larger than the value computed by DUSTY for our assumed grain type
and size range. Because the sub-mm observations are rather insensitive to emission
at radii beyond the half-power point of their 13\arcsec~beam, extending our
12\arcsec~diameter shell to larger sizes does not significantly enhance the model
850\micron~flux, and so does not help in resolving the problem of the sub-mm excess.
We find this expectation is supported from the results of new models in which we
have added an r$^{-2}$ envelope surrounding the r$^{-3}$ density distribution in
Model 7. We do not find any significant increase in the sub-mm wave flux density for
models with relative thicknesses of 15 and 30 for the outer shell
(Fig.\,\ref{mod-outshell}); moreover, in these cases the far-infrared modelled flux
begins to exceed the observed flux.

It is rather unlikely that the excess sub-mm flux can be explained by the
presence of small grains at radii inside of the half-power point of the
millimeter-wave beam, which are sufficiently cool (i.e., for providing the
excess observed submillimeter-wave flux) because they are located in
regions shielded from the direct stellar radiation. Such regions with
significantly higher-than-average extinction could plausibly exist in a
3-D physical model of I\,19475. However, the fundamental difficulty with
such a hypothesis is that, since small grains are such poor emitters at
sub-millimeter wavelengths, a very large mass of these grains is required
to produce the observed submillimeter emission -- and such a mass
contributes substantially at far-infrared wavelengths (i.e., $\sim$100
$\mu$m) for which the warm, small grains in our model already produce
adequate emission. For example, 0.01(20\,K/T$_d$)\ms~of cold dust at
temperature T$_d$ would be needed to produce the "excess" 850 $\mu$m flux
of 20 mJy, assuming $\kappa_{850} \sim$3 cm$^{2}$ gm$^{-1}$ (appropriate
for small grains; Jura, Webb \& Kahane 2001), but this dust contributes
far too much emission at far-infrared wavelengths (e.g. $\sim$20 Jy at
100$\mu$m) to be consistent with the data.

Although the nebula clearly has non-spherical components, simply
incorporating these by using a 3-D model, but without the addition of
large, cooler, dust grains, will not help resolve the problem of the
sub-mm excess. We conclude that it is likely that there are much larger
dust grains present within our model dust shell which are not accounted
for by our adopted dust grain distribution (which follows the standard MRN
prescription). Multi-wavelength observations of I\,19475's continuum flux
at radio, millimeter and sub-millimeter wavelengths will be needed in
order to probe the mass and sizes of these grains, and justify further
quantitative modelling with modified dust grain properties. High spatial
resolution submillimeter-wave interferometric observations with a facility
like the Submillimeter Array could help us identify the location of these
large, cool grains. One possibility is that the cool grains are associated with the
more compact ($\sim$5$''$) lobe-pair of this quadrupolar nebula as seen in
the HST image (Sahai 2004, Fig. 1).

Gledhill et al. (2002) analysed their 850$\mu$m data of I\,19475 (and other
sources), together with the 100$\mu$ flux, assuming a fixed dust
temperature of 120\,K. Although their data provide a potentially important
constraint for determining the dust content and mass in PPNs, their
analysis and its results (dust mass, dust emissivity power-law index) are
most likely invalid because we find from our modelling that the dust
temperatures are much lower, and cover a large range (46-94\,K). Gledhill
et al.'s derived dust mass, scaled to our adopted distance and multiplied
by our adopted gas-to-dust ratio, is 0.7\ms--that this number is not too
different from our estimate of $\gtrsim$1\ms, is purely coincidental. Our
modelling underscores the importance of first determining the mass and
temperature distribution of dust in PPNs carefully by fitting their full
near-to-far infrared SEDs, before inferring dust masses from
longer-wavelength data.

There are two other wavelength regions where the modelled flux is
discrepant from the data. First, in the $\sim (7-11) \mu$m range, the
model flux is lower than the observed PHT-S data. In this wavelength
region, the SED is at a minimum because both the stellar and the dust
shell contributions are rapidly declining at wavelengths approaching this
region. It is likely that the observed excess flux in this region results
from the presence of a small amount of warm dust (i.e. at temperatures
larger than those in our model dust shell -- e.g, 94--46\,K in the
best-fit model), and thus at radii smaller than the inner radius of the
cool dust shell. Second, in the $\sim (27-42) \mu$m range, the model flux
is larger than observed, suggesting a small excess of dust in our model
shell at some intermediate temperature range within the full range of
temperatures in our model shell. As we have discussed in Sec. 3.1, we
believe that amorphous water-ice is not a significant contributor in this
region and therefore it is unlikely to help in removing this discrepancy.
Both discrepancies are probably best addressed by going to a 2-D model of
the object.

\subsection{Modelling Uncertainties}

>From the off-source LWS spectrum (Fig.\, \ref{lwsspec}b), it is obvious that there is
an increasing and substantial amount of background contribution to the LWS source
data (Fig.\, \ref{lwsspec}a) beyond 90 $\mu$m. Although considerable care was taken
in subtracting the background spectrum from the source spectrum (Fig.\,
\ref{lwsspec}c), this remains a possible source of uncertainty in our modelling --
however, we do not think it affects our major conclusions. The need to adopt an
r$^{-3}$ (rather than a r$^{-2}$) density law is driven by a {\it systematic} excess
in the model flux above the observed flux over a large wavelength range, i.e. for
all wavelengths longwards of $\sim$70 $\mu$m (\S 4.4.2). This excess cannot be
purely a result of the uncertainty in the background correction, since the total
background correction upto 90 $\mu$m is less than $\sim$10\%, with the uncertainty in
the correction being significantly smaller. The submillimeter excess which we find in
I\,19475 is also too large to be accounted for by uncertainties in the LWS
background subtraction.

We would like to point out that in this paper, the emphasis in the fitting procedure
is not on the details of the mineralogy of the dust, but rather on the bulk dust
component which is comprised of amorphous silicates. Given that we do not have a
convincing detection of water-ice in the dust in I\,19475, a detailed investigation
of models with amorphous water-ice (e.g., its affect on the observed 43 $\mu$m
emission feature and the overall spectral shape) does not appear warranted.

\section{Conclusions}

We have reconstructed the spectral energy distribution of the pre-planetary
nebula IRAS 19475+3119, from the optical to the far-infrared, using ISO
spectra and broad-band photometric data. Features at 33.6 $\mu$m, 43
$\mu$m,, and possibly at 23 $\mu$m, identified in the ISO SWS spectrum of
the star, are most likely due to the presence of crystalline silicates.

The circumstellar dust envelope of the star was modelled using the 1-D
radiative transfer code, DUSTY. We find that the dust shell has a very
substantial mass, of $\gtrsim$1\ms, in all models which provide a
reasonable fit to the data. Our best fit is obtained for a shell with an
r$^{-3}$ radial density, inner and outer radii of 8.8$\times 10^{16}$ and
4.4 $\times 10^{17}$cm, dust temperatures ranging from about 94\,K to
46\,K, and $\tau$(0.55 $\mu$m)=1. The mass of this shell is
$\gtrsim$1\,[34 cm$^2$ g$^{-1}$/$\kappa(100\mu$m)][$\delta$/200]\ms, where
$\kappa$(100$\mu$m) is the 100$\mu$m~dust mass absorption coefficient (per
unit dust mass), and $\delta$ is the gas-to-dust ratio.

In agreement with results from optical imaging and millimeter-wave
observations of CO emission, our model fits support an r$^{-3}$ density
law for I\,19475's dust shell. The exponent of the density law
characterising the ambient circumstellar medium has important implications
for the interaction process between the fast collimated post-AGB winds and
the dense AGB envelopes which results in the observed shapes of PPNs and
PNs.

Our models show some discrepancies with the data. The most significant of
these is that the observed JCMT flux at sub-millimeter wavelengths
(850\micron) is a factor $\sim$2 larger than the model flux, suggesting
the presence of dust grains in the dust shell of I\,19475 which are
significantly larger than those accounted for by our adopted model dust
grain distribution.

RS is thankful to NASA for financial support for this study from the Astrophysics
Data Program (RTOP 399-20-00-08), and from the Space Telescope Science Institute
through Program number HST-GO-09463.01.  

\clearpage
\appendix

\section{Data reduction}

Offline processed SWS (OLP version 10.1), LWS (OLP version 10.0) and 
PHT-S (OLP version 10.0) data were retrieved from the ISO data archive. 
The SWS and LWS observations were further processed using 
ISAP (ISO Spectroscopic Analysis Package) version 2.1.  

\subsection {SWS} 

The data analysis using ISAP consisted of extensive
bad data removal primarily to minimize the effect of cosmic ray
hits. All detectors were compared to identify possible features.
For each sub-band, `offset' shifts were applied to bring  
the flux level of the 12 detectors to a mean value (Sturm 2000). 
The spectra of the 12 detectors were then averaged using median
clipping to discard points that lay more than 2.5$\sigma$ from 
the median flux. The averaging was done to a resolution of
300 and 1500 for SWS01 data taken with speed 1 and 4 respectively (Table
1).
The data of sub-band 3E(27.5$-$29.0 $\mu$m) are generally noisy
and unreliable (see e.g. Hrivnak et al. 2000; Hony et al. 2002). 
Our SWS data, below band 3A (16.5 $\mu$m) is very noisy.
This is also evident from the low IRAS flux at 12 $\mu$m (0.54 Jy). 
The averaged spectra of sub-band 3D (19.5$-$27.5 $\mu$m) were scaled 
in order to match the flux at 25 $\mu$m with the observed IRAS flux at 
this wavelength. Appropriate scaling factors were then applied to 
the remaining sub-bands in order to form a continuous spectrum from 
2.38$-$45.2 $\mu$m.

\subsection {LWS} 

Reduction of the LWS on and off-source observations consisted of 
extensive bad data removal using ISAP and rebinning on a fixed
resolution grid of $\lambda/\Delta\lambda$ = 250. In the case 
of the on-source data, for each detector, we examined the OLP dark current 
estimates made before (Dark 0) and after (Dark 1) the observation 
using LWS Interactive Analysis (LIA) version 10.2. To form a 
continuous spectrum from 43$-$197 $\mu$m, using ISAP\_SHIFT routine, we
applied 
a `DC offset' shift to the detectors which showed different Dark 0 and Dark
1
values. Dark current measurements of detectors LW3 (123$-$152 $\mu$m), 
LW4 (142$-$171 $\mu$m) and LW5 (161$-$197 $\mu$m) showed a decreasing 
temporal shift. Hence a `gain' correction was applied to these 3 detectors 
using the ISAP\_SHIFT routine (Molinari) to match the flux in the 
overlapping spectral regions between the detectors.

The background spectrum was rebinned to the same wavelength scale as the
source
spectrum. It was assumed that the [C~II] emission at 158 $\mu$m in the
on-source
spectrum is entirely interstellar in origin. The background spectrum was
therefore scaled in the region of the [C~II] line (LW4 detector) to match
the
line strength in the on and off-source data. Then, using the LW4 detector
as a
reference the remaining detectors were scaled to form a continuous
off-source
spectrum (Fig.\,\ref{lwsspec}b). Subsequently, the LWS spectrum of I\,19475
(Fig.\,\ref{lwsspec}c) was obtained by subtracting the off-source spectrum
(Fig.\,\ref{lwsspec}b) from the on-source data (Fig.\,\ref{lwsspec}a).

\subsection {PHT-S}

The fully processed PHT-S data of I\,19475 retrieved from the ISO
data archive is scientifically validated and was used directly (i.e. with
no
further
processing).  
The typical calibration accuracy for PHT-S observations made in 
staring, extended source mode is better than $\pm$15\% (Klaas et al.,
2002).
The PHT-S data, I$_{\lambda}$(i) in units of surface brightness 
(W m$^{-2}$ $\mu$m$^{-1}$ sr$^{-1}$) obtained from the  archive 
was converted to flux density, F$_{\lambda}$(i) units 
(W m$^{-2}$ $\mu$m$^{-1}$) using the receipe given in 
Laureijs et al. (2003) :

$$F_{\lambda}(i) = 10^{-6}I_{\lambda}(i)\frac{C^{e}_{ave}}{C^{p}_{ave}}$$

where, \\
\begin{itemize}
\item $I_{\lambda}$(i) in W m$^{-2}$ $\mu$m$^{-1}$ sr$^{-1}$ is the 
surface brightness for a given detector array element $i$;
\item $\lambda(i)$ in $\mu$m is the central wavelength of a PHT-S
detector pixel;
\item $C^{e}_{ave}(i)$ in (Vs$^{-1}$)/(MJysr$^{-1}$) is the
average spectral response function for the conversion from signal
in Vs$^{-1}$ to surface brightness in MJysr$^{-1}$;
\item $C^{p}_{ave}(i)$ in (Vs$^{-1}$)/Jy is the average spectral
response function for the conversion from signal in (Vs$^{-1}$)
to point source flux density in Jy.
\end{itemize}

The conversion factors $C^{e}_{ave}$ and $C^{p}_{ave}$ (Table 6)
were obtained from the PSPECAL.FITS file which is part of the 
general calibration files (Cal-G) distributed with the 
ISOPHOT Interactive Analysis (PIA), version 10.0.
\clearpage
\begin{table}
\begin{center}
\caption{Conversion factors for PHT-S data}
\begin{tabular}{|c|c|c|c|}
\hline
Wavelength & $C^{e}_{ave}(i)$ &  $C^{p}_{ave}(i)$ & {\Large
${C^{e}_{ave}(i)}
\over
{C^{p}_{ave}(i)}$}\\
($\mu$m)   &  (Vs$^{-1}$)/(MJysr$^{-1}$) & (Vs$^{-1}$)/Jy &
Jy/(MJysr$^{-1}$)\\
\hline \hline
      2.46870 &  3.03987 $\times 10^{-5}$ &  0.00333517 & 9.11459 $\times
10^{-3}$\\
      2.50950 &  3.01950 $\times 10^{-5}$ &  0.00328285 & 9.19780 $\times
10^{-3}$\\
      2.55010 &  3.23561 $\times 10^{-5}$ &  0.00348378 & 9.28764 $\times
10^{-3}$\\
      2.59070 &  3.14137 $\times 10^{-5}$ &  0.00335591 & 9.36071 $\times
10^{-3}$\\
      2.63110 &  3.55078 $\times 10^{-5}$ &  0.00376118 & 9.44060 $\times
10^{-3}$\\
      2.67150 &  3.47183 $\times 10^{-5}$ &  0.00364872 & 9.51520 $\times
10^{-3}$\\
      2.71180 &  3.33576 $\times 10^{-5}$ &  0.00348496 & 9.57187 $\times
10^{-3}$\\
      2.75210 &  3.41233 $\times 10^{-5}$ &  0.00354108 & 9.63641 $\times
10^{-3}$\\
      2.79220 &  3.62736 $\times 10^{-5}$ &  0.00374119 & 9.69574 $\times
10^{-3}$\\
      2.83230 &  3.82493 $\times 10^{-5}$ &  0.00391801 & 9.76243 $\times
10^{-3}$\\
      2.87220 &  4.02213 $\times 10^{-5}$ &  0.00409953 & 9.81120 $\times
10^{-3}$\\
      2.91210 &  3.68381 $\times 10^{-5}$ &  0.00373792 & 9.85524 $\times
10^{-3}$\\
      2.95190 &  4.81234 $\times 10^{-5}$ &  0.00485787 & 9.90628 $\times
10^{-3}$\\
      2.99170 &  5.82109 $\times 10^{-5}$ &  0.00585568 & 9.94093 $\times
10^{-3}$\\
      3.03130 &  7.06671 $\times 10^{-5}$ &  0.00708768 & 9.97041 $\times
10^{-3}$\\
      3.07090 &  6.49100 $\times 10^{-5}$ &  0.00648637 & 1.00071 $\times
10^{-2}$\\
      3.11030 &  6.87588 $\times 10^{-5}$ &  0.00684926 & 1.00389 $\times
10^{-2}$\\
      3.14970 &  6.13806 $\times 10^{-5}$ &  0.00610456 & 1.00549 $\times
10^{-2}$\\
      3.18900 &  7.19383 $\times 10^{-5}$ &  0.00713846 & 1.00776 $\times
10^{-2}$\\
      3.22830 &  6.93928 $\times 10^{-5}$ &  0.00687342 & 1.00958 $\times
10^{-2}$\\
      3.26740 &  7.03575 $\times 10^{-5}$ &  0.00695950 & 1.01096 $\times
10^{-2}$\\
      3.30650 &  6.28810 $\times 10^{-5}$ &  0.00621426 & 1.01188 $\times
10^{-2}$\\
      3.34550 &  6.44410 $\times 10^{-5}$ &  0.00636565 & 1.01232 $\times
10^{-2}$\\
      3.38430 &  5.97321 $\times 10^{-5}$ &  0.00590055 & 1.01231 $\times
10^{-2}$\\
      3.42320 &  6.31887 $\times 10^{-5}$ &  0.00624471 & 1.01188 $\times
10^{-2}$\\
      3.46190 &  5.41975 $\times 10^{-5}$ &  0.00536089 & 1.01098 $\times
10^{-2}$\\
      3.50050 &  5.34569 $\times 10^{-5}$ &  0.00529456 & 1.00966 $\times
10^{-2}$\\
\hline
\end{tabular}
\end{center}
\end{table}

\setcounter{table}{5}
\begin{table}
\begin{center}
\caption{contd....}
\begin{tabular}{|c|c|c|c|}
\hline
Wavelength & $C^{e}_{ave}(i)$ &  $C^{p}_{ave}(i)$ & {\Large
${C^{e}_{ave}(i)}
\over
{C^{p}_{ave}(i)}$}\\
($\mu$m)   &  (Vs$^{-1}$)/(MJysr$^{-1}$) & (Vs$^{-1}$)/Jy &
Jy/(MJysr$^{-1}$)\\
\hline \hline
      3.53910 &  4.24939$\times 10^{-5}$  &  0.00421167 & 1.00896 $\times
10^{-2}$\\
      3.57760 &  5.09418$\times 10^{-5}$  &  0.00506002 & 1.00675 $\times
10^{-2}$\\
      3.61600 &  4.31928$\times 10^{-5}$  &  0.00430166 & 1.00410 $\times
10^{-2}$\\
      3.65430 &  3.78045$\times 10^{-5}$  &  0.00377251 & 1.00210 $\times
10^{-2}$\\
      3.69250 &  2.49969$\times 10^{-5}$  &  0.00250331 & 9.98554 $\times
10^{-3}$\\
      3.73070 &  3.35651$\times 10^{-5}$  &  0.00329863 & 1.01755 $\times
10^{-2}$\\
      3.76870 &  3.71818$\times 10^{-5}$  &  0.00364488 & 1.02011 $\times
10^{-2}$\\
      3.80670 &  3.82232$\times 10^{-5}$  &  0.00373432 & 1.02357 $\times
10^{-2}$\\
      3.84460 &  3.29000$\times 10^{-5}$  &  0.00320791 & 1.02559 $\times
10^{-2}$\\
      3.88240 &  3.63065$\times 10^{-5}$  &  0.00352998 & 1.02852 $\times
10^{-2}$\\
      3.92010 &  3.44827$\times 10^{-5}$  &  0.00334782 & 1.03000 $\times
10^{-2}$\\
      3.95780 &  3.76135$\times 10^{-5}$  &  0.00364328 & 1.03241 $\times
10^{-2}$\\
      3.99540 &  3.25984$\times 10^{-5}$  &  0.00315459 & 1.03336 $\times
10^{-2}$\\
      4.03280 &  3.26180$\times 10^{-5}$  &  0.00315082 & 1.03522 $\times
10^{-2}$\\
      4.07020 &  2.89407$\times 10^{-5}$  &  0.00279106 & 1.03691 $\times
10^{-2}$\\
      4.10760 &  3.27721$\times 10^{-5}$  &  0.00315620 & 1.03834 $\times
10^{-2}$\\
      4.14480 &  2.82243$\times 10^{-5}$  &  0.00271870 & 1.03815 $\times
10^{-2}$\\
      4.18190 &  3.10457$\times 10^{-5}$  &  0.00298790 & 1.03905 $\times
10^{-2}$\\
      4.21900 &  2.72054$\times 10^{-5}$  &  0.00261666 & 1.03970 $\times
10^{-2}$\\
      4.25600 &  2.69869$\times 10^{-5}$  &  0.00259448 & 1.04017 $\times
10^{-2}$\\
      4.29290 &  2.27141$\times 10^{-5}$  &  0.00218314 & 1.04043 $\times
10^{-2}$\\
      4.32970 &  2.39768$\times 10^{-5}$  &  0.00230451 & 1.04043 $\times
10^{-2}$\\
      4.36640 &  2.07467$\times 10^{-5}$  &  0.00199824 & 1.03825 $\times
10^{-2}$\\
      4.40310 &  2.35388$\times 10^{-5}$  &  0.00226821 & 1.03777 $\times
10^{-2}$\\
      4.43970 &  1.94929$\times 10^{-5}$  &  0.00187972 & 1.03701 $\times
10^{-2}$\\
      4.47610 &  2.18277$\times 10^{-5}$  &  0.00210694 & 1.03599 $\times
10^{-2}$\\
      4.51250 &  1.81484$\times 10^{-5}$  &  0.00175392 & 1.03473 $\times
10^{-2}$\\
\hline
\end{tabular}
\end{center}
\end{table}

\setcounter{table}{5}
\begin{table}
\begin{center}
\caption{contd....}
\begin{tabular}{|c|c|c|c|}
\hline
Wavelength & $C^{e}_{ave}(i)$ &  $C^{p}_{ave}(i)$ & {\Large
${C^{e}_{ave}(i)}
\over
{C^{p}_{ave}(i)}$}\\
($\mu$m)   &  (Vs$^{-1}$)/(MJysr$^{-1}$) & (Vs$^{-1}$)/Jy &
Jy/(MJysr$^{-1}$) \\
\hline \hline
      4.54890 &  1.98181$\times 10^{-5}$  &  0.00191809 & 1.03322 $\times
10^{-2}$\\
      4.58510 &  1.62796$\times 10^{-5}$  &  0.00157838 & 1.03141 $\times
10^{-2}$\\
      4.62120 &  1.42384$\times 10^{-5}$  &  0.00138326 & 1.02934 $\times
10^{-2}$\\
      4.65730 &  1.44786$\times 10^{-5}$  &  0.00140976 & 1.02703 $\times
10^{-2}$\\
      4.69330 &  1.50828$\times 10^{-5}$  &  0.00147225 & 1.02447 $\times
10^{-2}$\\
      4.72920 &  1.36050$\times 10^{-5}$  &  0.00132851 & 1.02408 $\times
10^{-2}$\\
      4.76500 &  1.80055$\times 10^{-5}$  &  0.00176359 & 1.02096 $\times
10^{-2}$\\
      4.80080 &  1.34228$\times 10^{-5}$  &  0.00131913 & 1.01755 $\times
10^{-2}$\\
      4.83640 &  1.44308$\times 10^{-5}$  &  0.00142338 & 1.01384 $\times
10^{-2}$\\
      5.83960 &  12.8575$\times 10^{-5}$  &   0.0115198 & 1.11612 $\times
10^{-2}$\\
      5.93380 &  11.4284$\times 10^{-5}$  &   0.0102716 & 1.11262 $\times
10^{-2}$\\
      6.02800 &  11.2309$\times 10^{-5}$  &   0.0100773 & 1.11448 $\times
10^{-2}$\\
      6.12200 &  11.6475$\times 10^{-5}$  &   0.0104899 & 1.11035 $\times
10^{-2}$\\
      6.21600 &  10.9303$\times 10^{-5}$  &  0.00983196 & 1.11171 $\times
10^{-2}$\\
      6.30990 &  8.64221$\times 10^{-5}$  &  0.00776619 & 1.11280 $\times
10^{-2}$\\
      6.40380 &  10.6843$\times 10^{-5}$  &  0.00964404 & 1.10787 $\times
10^{-2}$\\
      6.49750 &  9.88445$\times 10^{-5}$  &  0.00891797 & 1.10837 $\times
10^{-2}$\\
      6.59120 &  9.40491$\times 10^{-5}$  &  0.00848214 & 1.10879 $\times
10^{-2}$\\
      6.68480 &  8.12382$\times 10^{-5}$  &  0.00732521 & 1.10902 $\times
10^{-2}$\\
      6.77830 &  9.59879$\times 10^{-5}$  &  0.00870225 & 1.10302 $\times
10^{-2}$\\
      6.87170 &  9.17306$\times 10^{-5}$  &  0.00831815 & 1.10278 $\times
10^{-2}$\\
      6.96500 &  10.4062$\times 10^{-5}$  &  0.00944026 & 1.10232 $\times
10^{-2}$\\
      7.05830 &  9.92587$\times 10^{-5}$  &  0.00900942 & 1.10172 $\times
10^{-2}$\\
      7.15150 &  9.66050$\times 10^{-5}$  &  0.00877220 & 1.10126 $\times
10^{-2}$\\
      7.24460 &  8.08621$\times 10^{-5}$  &  0.00734715 & 1.10059 $\times
10^{-2}$\\
      7.33760 &  7.66112$\times 10^{-5}$  &  0.00696665 & 1.09968 $\times
10^{-2}$\\
      7.43060 &  6.90864$\times 10^{-5}$  &  0.00628884 & 1.09856 $\times
10^{-2}$\\
\hline
\end{tabular}
\end{center}
\end{table}

\setcounter{table}{5}
\begin{table}
\begin{center}
\caption{contd....}
\begin{tabular}{|c|c|c|c|}
\hline
Wavelength & $C^{e}_{ave}(i)$ &  $C^{p}_{ave}(i)$ & {\Large
${C^{e}_{ave}(i)}
\over
{C^{p}_{ave}(i)}$}\\
($\mu$m)   &  (Vs$^{-1}$)/(MJysr$^{-1}$) & (Vs$^{-1}$)/Jy &
Jy/(MJysr$^{-1}$)\\
\hline \hline
      7.52350 &  8.15585$\times 10^{-5}$  &  0.00743328 & 1.09721 $\times
10^{-2}$\\
      7.61630 &  7.35407$\times 10^{-5}$  &  0.00671185 & 1.09568 $\times
10^{-2}$\\
      7.70900 &  7.07851$\times 10^{-5}$  &  0.00647062 & 1.09395 $\times
10^{-2}$\\
      7.80160 &  6.83308$\times 10^{-5}$  &  0.00625740 & 1.09200 $\times
10^{-2}$\\
      7.89420 &  7.86450$\times 10^{-5}$  &  0.00716019 & 1.09836 $\times
10^{-2}$\\
      7.98660 &  7.69969$\times 10^{-5}$  &  0.00702496 & 1.09605 $\times
10^{-2}$\\
      8.07900 &  7.75643$\times 10^{-5}$  &  0.00709291 & 1.09355 $\times
10^{-2}$\\
      8.17140 &  6.57546$\times 10^{-5}$  &  0.00602763 & 1.09089 $\times
10^{-2}$\\
      8.26360 &  6.08753$\times 10^{-5}$  &  0.00559467 & 1.08809 $\times
10^{-2}$\\
      8.35580 &  5.63152$\times 10^{-5}$  &  0.00514833 & 1.09385 $\times
10^{-2}$\\
      8.44780 &  6.16770$\times 10^{-5}$  &  0.00565487 & 1.09069 $\times
10^{-2}$\\
      8.53980 &  5.88137$\times 10^{-5}$  &  0.00540839 & 1.08745 $\times
10^{-2}$\\
      8.63180 &  6.33101$\times 10^{-5}$  &  0.00583982 & 1.08411 $\times
10^{-2}$\\
      8.72360 &  3.26216$\times 10^{-5}$  &  0.00299513 & 1.08915 $\times
10^{-2}$\\
      8.81540 &  5.01120$\times 10^{-5}$  &  0.00490196 & 1.02228 $\times
10^{-2}$\\
      8.90700 &  5.79664$\times 10^{-5}$  &  0.00562173 & 1.03111 $\times
10^{-2}$\\
      8.99860 &  5.44365$\times 10^{-5}$  &  0.00519205 & 1.04846 $\times
10^{-2}$\\
      9.09020 &  5.10184$\times 10^{-5}$  &  0.00483174 & 1.05590 $\times
10^{-2}$\\
      9.18160 &  4.59752$\times 10^{-5}$  &  0.00428244 & 1.07357 $\times
10^{-2}$\\
      9.27300 &  4.28071$\times 10^{-5}$  &  0.00396460 & 1.07973 $\times
10^{-2}$\\
      9.36430 &  4.05319$\times 10^{-5}$  &  0.00373380 & 1.08554 $\times
10^{-2}$\\
      9.45550 &  4.54677$\times 10^{-5}$  &  0.00412516 & 1.10229 $\times
10^{-2}$\\
      9.54660 &  4.58967$\times 10^{-5}$  &  0.00414712 & 1.10671 $\times
10^{-2}$\\
      9.63760 &  3.91220$\times 10^{-5}$  &  0.00348406 & 1.12289 $\times
10^{-2}$\\
      9.72860 &  3.55302$\times 10^{-5}$  &  0.00315427 & 1.12642 $\times
10^{-2}$\\
      9.81950 &  2.83132$\times 10^{-5}$  &  0.00247981 & 1.14175 $\times
10^{-2}$\\
      9.91030 &  3.39671$\times 10^{-5}$  &  0.00296851 & 1.14425 $\times
10^{-2}$\\
\hline
\end{tabular}
\end{center}
\end{table}

\setcounter{table}{5}
\begin{table}
\begin{center}
\caption{contd....}
\begin{tabular}{|c|c|c|c|}
\hline
Wavelength & $C^{e}_{ave}(i)$ &  $C^{p}_{ave}(i)$ & {\Large
${C^{e}_{ave}(i)}
\over
{C^{p}_{ave}(i)}$}\\
($\mu$m)   &  (Vs$^{-1}$)/(MJysr$^{-1}$) & (Vs$^{-1}$)/Jy &
Jy/(MJysr$^{-1}$)\\
\hline \hline
      10.0010 &  3.34585$\times 10^{-5}$  &  0.00288679 & 1.15902 $\times
10^{-2}$\\
      10.0917 &  2.71791$\times 10^{-5}$  &  0.00234248 & 1.16027 $\times
10^{-2}$\\
      10.1823 &  2.93380$\times 10^{-5}$  &  0.00249772 & 1.17459 $\times
10^{-2}$\\
      10.2727 &  3.21568$\times 10^{-5}$  &  0.00273768 & 1.17460 $\times
10^{-2}$\\
      10.3632 &  2.83052$\times 10^{-5}$  &  0.00238194 & 1.18833 $\times
10^{-2}$\\
      10.4535 &  2.61391$\times 10^{-5}$  &  0.00220172 & 1.18721 $\times
10^{-2}$\\
      10.5437 &  2.52504$\times 10^{-5}$  &  0.00210409 & 1.20006 $\times
10^{-2}$\\
      10.6339 &  2.60993$\times 10^{-5}$  &  0.00217871 & 1.19792 $\times
10^{-2}$\\
      10.7240 &  2.67050$\times 10^{-5}$  &  0.00220733 & 1.20983 $\times
10^{-2}$\\
      10.8140 &  2.50139$\times 10^{-5}$  &  0.00207279 & 1.20677 $\times
10^{-2}$\\
      10.9040 &  2.50909$\times 10^{-5}$  &  0.00206055 & 1.21768 $\times
10^{-2}$\\
      10.9938 &  2.42555$\times 10^{-5}$  &  0.00199853 & 1.21367 $\times
10^{-2}$\\
      11.0836 &  2.15947$\times 10^{-5}$  &  0.00176475 & 1.22367 $\times
10^{-2}$\\
      11.1733 &  2.05535$\times 10^{-5}$  &  0.00166620 & 1.23356 $\times
10^{-2}$\\
      11.2629 &  1.76285$\times 10^{-5}$  &  0.00143567 & 1.22789 $\times
10^{-2}$\\
      11.3524 &  1.74882$\times 10^{-5}$  &  0.00141409 & 1.23671 $\times
10^{-2}$\\
      11.4419 &  2.05966$\times 10^{-5}$  &  0.00167428 & 1.23018 $\times
10^{-2}$\\
      11.5313 &  2.31557$\times 10^{-5}$  &  0.00187026 & 1.23810 $\times
10^{-2}$\\
      11.6206 &  1.98068$\times 10^{-5}$  &  0.00159024 & 1.24552 $\times
10^{-2}$\\
\hline
\end{tabular}
\end{center}
\end{table}

\end{document}